\documentclass[twocolumn]{aastex631}
\usepackage{soul}
\usepackage{booktabs}

\begin{document}

\title{Large Scale Overdensity of Lyman Break Galaxies Around the $z=6.3$ Ultraluminous Quasar J0100+2802}

\correspondingauthor{Maria Pudoka}
\author[0000-0003-4924-5941]{Maria Pudoka}
\affiliation{Steward Observatory, University of Arizona, 933 North Cherry Avenue, Tucson, AZ 85721-0065, USA}
\email{pudoka@arizona.edu}

\author[0000-0002-7633-431X]{Feige Wang}
\affiliation{Steward Observatory, University of Arizona, 933 North Cherry Avenue, Tucson, AZ 85721-0065, USA}

\author[0000-0003-3310-0131]{Xiaohui Fan}
\affiliation{Steward Observatory, University of Arizona, 933 North Cherry Avenue, Tucson, AZ 85721-0065, USA}

\author[0000-0001-5287-4242]{Jinyi Yang}
\affiliation{Steward Observatory, University of Arizona, 933 North Cherry Avenue, Tucson, AZ 85721-0065, USA}

\author[0000-0002-6184-9097]{Jaclyn Champagne}
\affiliation{Steward Observatory, University of Arizona, 933 North Cherry Avenue, Tucson, AZ 85721-0065, USA}

\author{Victoria Jones}
\affiliation{Steward Observatory, University of Arizona, 933 North Cherry Avenue, Tucson, AZ 85721-0065, USA}

\author[0000-0002-1620-0897]{Fuyan Bian}
\affiliation{European Southern Observatory, Alonso de Cordova 3107, Casilla 19001, Vitacura, Santiago 19, Chile}

\author[0000-0001-8467-6478]{Zheng Cai}
\affiliation{Department of Astronomy, Tsinghua University, Beijing 100084, People's Republic of China}

\author[0000-0003-4176-6486]{Linhua Jiang}
\affiliation{Department of Astronomy, School of Physics, Peking University, Beijing 100871, People's Republic of China}
\affiliation{Kavli Institute for Astronomy and Astrophysics, Peking University, Beijing 100871, People's Republic of China}

\author[0000-0002-0409-5719]{Dezi Liu}
\affiliation{South-Western Institute for Astronomy Research, Yunnan University, Kunming, 650500, People's Republic of China}


\author[0000-0002-7350-6913]{Xue-Bing Wu}
\affiliation{Department of Astronomy, School of Physics, Peking University, Beijing 100871, People's Republic of China}
\affiliation{Kavli Institute for Astronomy and Astrophysics, Peking University, Beijing 100871, People's Republic of China}


\begin{abstract}
We study the environment of the  $z=6.33$ ultraluminous quasar SDSS J010013.02+280225.8 (J0100) to understand its association with large-scale structure.  Theoretical models propose high-redshift quasars as markers of galaxy overdensities residing in the most massive dark matter halos (DMHs) in the early universe.  J0100 is an ultraluminous quasar with the most massive black hole known at $z\gtrsim6$, suggesting a high likelihood of residing in a massive DMH.  We present wide-field ($\sim522$~square arcminute) imaging in the $r$-, $i$-, and $z$-bands from the Large Binocular Camera on the Large Binocular Telescope, with $Y$- and $J$-band imaging from the Wide-field Infrared Camera on the Canada-France-Hawaii Telescope, centered on J0100. Applying color selections, we identify 23 objects as $i$-droput Lyman Break Galaxy (LBG) candidates in the J0100 field. We use the deep photometric catalog in the 1.27 square degree COSMOS field to calculate the density of LBGs in a blank field, and to estimate the selection completeness and purity. 
The observed surface density of LBG candidates in the J0100 field corresponds to a galaxy overdensity of $\delta=4$ (at 8.4$\sigma$). This large-scale overdensity suggests that the $\sim 22$ square arcminute overdensity found by Kashino et al. using JWST data extends out to much larger scales. We calculate the angular auto-correlation function of the candidates and find a positive correlation on $\lesssim 10$ arcminute scales as well as evidence of asymmetries in their spatial distribution, further suggesting a direct detection of large-scale structure around the ultra-luminous quasar J0100.
\end{abstract}

\section{Introduction} \label{sec:intro}


Over the last two decades of quasar research, ground-based surveys have unveiled the existence of a large population of luminous quasars at $z\sim6$; residing well within the epoch of reionization (EoR), these quasars have black hole masses $\geq 10^{9}\,M_{\odot}$ \citep{venemans, wu, banados16, jiang, yang, wang19, yang21, fan23}.  These rare quasars, powered by such massive black holes, require that the black holes must have grown to their current state in less than 1 Gyr after the Big Bang.  Their formation and subsequent growth in such a short period of time have provoked theoretical exploration into many possible evolutionary scenarios for supermassive black holes \citep[SMBHs;][]{inayoshi, volonteri}.  

Cosmological simulations can produce these SMBHs by $z\sim 6$ by allowing exceptionally high accretion rates (super-Eddington) or starting with massive ($<10^{3-4}\,M_{\odot}$) seeds.  These simulations \citep{springel, overzier09} along with the highly clustered nature of quasars \citep{mow, eft}, and quasar abundance matching \citep{lukic} all indicate that these quasars reside in the most massive dark matter halos (DMHs).  SMBHs then grow through two essential processes: 1) accreting cold gas and 2) merging with other black holes following the idea of hierarchical structure formation \citep{haehnelt, dimatteo05, dimatteo12}.  The scenarios framing the formation and growth of these quasars suggest that they reside in overdense environments of galaxies as they must 1) reside in the most massive DMHs that typically host clusters, 2) be surrounded by large reservoirs of gas from  which they can accrete, and 3) be near many other black holes with which they will merge.  In the most extreme overdense regions, these galaxy overdensties could eventually settle into galaxy clusters with $M\sim10^{14-15}\,M_{\odot}$ by the present day \citep{costa14}. The progenitors of these are known as protoclusters.

Galaxy clusters and protoclusters play a significant role in advancing our understanding of the formation and evolution of the universe.  Protoclusters, in particular, provide valuable insights into the growth of early structure formation. The distribution of DMHs on a cosmic scale is theoretically traced by luminous matter: galaxies and protoclusters at high redshifts \citep{adelberger}. Comparing the observed structures and properties of protoclusters at high redshifts to cosmological simulations can also help to test various theories for dark matter or cosmological initial conditions \citep[see][for review]{overzier16}.  Additionally, the high star formation rates in these early structures \citep{costa14,chiang17} likely played a role in carving out bubbles of ionized hydrogen during the EoR, opening a window to constrain the ionizing radiation field of early galaxies and the topology of reionization \citep{lily}.  Furthermore, probing the galaxy properties in these dense environments at such early stages can shed light on galaxy formation and evolution, particularly how this differs from galaxy formation and evolution in more typical, less dense, regions \citep{nantais, lb}.  
 
Though theory predicts overdense regions around high redshift quasars \citep{overzier09, romano}, there have been mixed results when it comes to observations of the environments of these quasars \citep{kim09}.  Some authors have reported overdensities of galaxies in the quasar environments \citep{kashikawa, utsumi, balmaverde} while others have found no significant evidence of an overdensity \citep{ mazzucchelli, banados, willot}, and in some cases, even  underdense environments have been reported \citep{simpson}.  Many hypotheses have been proposed to account for these inconsistencies.  As discussed below, these include small fields of view (FoVs), strong quasar feedback, and differing selection techniques.

Overdensities anchored by quasars at $z\sim 6$ should easily extend to several tens of comoving Mpc (cMpc) away from the central quasar \citep{overzier09, chiang13}.  These distances correspond to fields of view as large as $\sim30'\times30'$ at $z\sim 6$.  Many searches use deep imagers with FoVs on the order of only a few arcminutes on a side corresponding to $\lesssim 10$ cMpc at $z\sim 6$ \citep{stiavelli, kim09, simpson, mazzucchelli}.  Using these small FoVs can lead to missing many galaxies that are part of the structure, thus diluting the overdensity signal.  Another plausible explanation includes powerful quasar feedback heating the intergalactic medium (IGM) on scales up to a few cMpc \citep{babul, scannapieco}. This ionizing radiation can prevent star formation (at least in the lowest mass galaxies) and reduce the ability to observe Ly$\alpha$ emitters (LAEs) or Lyman Break Galaxies (LBGs) tracing the dark matter overdensities near the quasar \citep[e.g.,][]{utsumi}.  It is evident that varying sizes of the FoV can severely affect the detection of an overdensity \citep{chiang13}.
 
Another difficulty with drawing conclusions based on the results of these studies arises from the fact that various groups use different selection techniques.  Some authors search for submillimeter galaxies (SMGs) \citep{champagne, meyer} or [\ion{O}{3}] emitters \citep{kashino, wang23}, while others look for LBGs or LAEs \citep{banados, morselli, balmaverde, mignoli, champgne23}.  Some fields even show conflicting results depending on which type of galaxy is selected \citep[e.g.,][]{ota}.  This occurred also in \citet{utsumi} and \citet{goto} in which an overdensity of LBGs was initially detected and a follow-up search for LAEs resulted in no overdensity.  It is clear that the search for dark matter overdensities traced by biased galaxy populations is heavily reliant on both observational constraints and the chosen galaxy selection. 

With the launch of JWST, there have been a number of efforts within the past year to search for [\ion{O}{3}] emitters in the fields of massive quasars using JWST's deep NIRCam wide-field slitless spectroscopic capabilities.  Though the FoV is small (two $2.2'\times2.2'$ detectors), the initial findings of these probes into $z\sim6$ quasar environments have found many instances of galaxy overdensities.  For example, \citet{kashino} surveyed a $6.5'\times3.4'$ area around J0100+2802 and found 24 [\ion{O}{3}] emitting systems exactly at the redshift of the quasar, many more than those at foreground redshifts from the quasar.  Additionally, \citet{wang23}, discovered a filamentary structure consisting of the luminous quasar J0305–3150 and 10 [\ion{O}{3}] emitters at $z = 6.6$ making this an overdensity of $\delta = 12.6$ in a FoV of one NIRCam pointing.  The strength of JWST in probing the faint end of the galaxy overdensity is evident, but ground-based searches for bright galaxies over much wider fields are still valuable to detect the full spatial extent of the overdensities.

In this study, we analyze the $\sim58\times58$ cMpc$^2$ (or $\sim$~$8\times8$ physical Mpc$^2$ at $z=6.33$) field around the ultraluminous quasar J010013.02+280225.8 (J0100).  J0100 is the most luminous quasar powered by the most massive black hole identified at $z\gtrsim6$ \citep{wu}.  With a luminosity of $L_{bol}\sim10^{48}$ erg s$^{-1}$ \citep{wu}, a black hole mass of $M_{BH}\sim10^{10}$ M$_{\odot}$ \citep{eilers}, and a host galaxy mass of $M_{\rm{dyn}}\geq 7\times 10^{10}\,M_{\odot}$ \citep{wang19a}, it is an ideal candidate to reside in a massive dark matter halo capable of hosting a galaxy overdensity.  Our analysis uses data from the Large Binocular Camera (LBC) on the Large Binocular Telescope (LBT) and the Wide-field Infrared Camera (WIRCam) on the Canada-France-Hawaii Telescope (CFHT).  These instruments provide a simultaneously wide and deep optical and near-infrared (NIR) imaging of the quasar field with a FoV of $\sim 25'\times23'$ in the $r$-, $i$-, $z$-, $Y$-, and $J$- filters. 

As mentioned, an overdensity of [\ion{O}{3}] emitters has been detected within an area of $6.5'\times3.4'$ centered on J0100 using JWST/NIRCam slitless spectroscopy \citep{kashino}.  While this spatial scale probes out to roughly $7$ cMpc away from the quasar, we alternatively focus on selecting LBGs as tracers of the large-scale structure of dark matter on scales up to $\sim$25 cMpc away from the quasar.  Our objective is to investigate the large-scale environment at protocluster scales using the photometric $i$-dropout technique discussed in Section~\ref{sec:selection}.

The structure of this paper is as follows: In section~\ref{sec:obs}, we describe the observations, data reduction, and the photometric catalog used for the subsequent analysis.  Section~\ref{sec:selection} discusses the selection criteria for the LBG candidates including filtering out contaminants.  Section~\ref{expect} describes the calculation of the expected number of LBGs in a blank field and the completeness/contamination of the sample.  In section~\ref{sec:overdensity}, we present the evidence for the existence of a galaxy overdensity and examine the spatial distribution of the high-redshift candidates.  Finally, in section~\ref{sec:summary}, we summarize our findings.  All magnitudes are reported in the AB system and we adopt a $\Lambda$CDM cosmology in which $H_0 = 70\,\mathrm{km\,s^{-1}\,Mpc^{-1}}$, $\Omega_m = 0.3$, and $\Omega_{\Lambda} = 0.7$ in which $1''= 40.6$ ckpc at $z=6.3$.

\section{Observations and Data Reduction} \label{sec:obs}
\subsection{LBT and CFHT Observations}
For this analysis, observations were taken using the Large Binocular Cameras (LBCs) on the Large Binocular Telescope (LBT) along with additional data from the Wide-field Infrared Camera (WIRCam) on the Canada-France-Hawaii Telescope (CFHT).  The observation designs are described below. 

The LBCs are two wide-field imagers mounted on the prime focus of the LBT. The LBC Blue is optimized for observations from 3500 \AA\ to 6500 \AA, while the LBC Red is optimized for observations from 5500 \AA\ to 1 micron. Both cameras have field of view (FoV) of $\sim25'\times23'$ \citep{Giallongo,speziali}. 
The LBC observations were obtained on 2015 November 22 (UT) under clear conditions (PI: X. Fan). To enable the $i$-dropout selection, we observed the J0100 field with $i_{\rm SDSS}$ and $z_{\rm SDSS}$ filters on the red channel of LBC. Additionally, we obtained $r_{\rm SDSS}$ imaging on the blue side of LBC simultaneously. The individual exposures for all images were set to 100 seconds to minimize the effects from cosmic rays and the saturation of bright stars in the field. The total on-source exposure in $r$-band, $i$-band, and $z$-band are 3.7 hours, 1.5 hours, and 2.1 hours, respectively. 


Furthermore, we performed NIR imaging with WIRCam on CFHT for J0100 (PI: J. Yang, RunID: 17BS03). The WIRCam is a wide field imager and has a FoV of $\sim21.5'\times21.5'$ \citep{puget}. Taking advantage of the large FoV, the WIRCam observations could fully cover the FoV of LBC with a carefully designed dithering pattern. Following \cite{balmaverde}, we selected the $Y$ and $J$ broad band filters for this program. The data have been obtained in Queue mode through the Telescope Access Program of NAOC\footnote{\url{https://tap.china-vo.org}} during the 2017B semester. The individual exposures for the $Y$-band and $J$-band were 120 seconds and 60 seconds, respectively. In total, we integrated 6.0 hours and 6.9 hours for $Y$-band and $J$-band, respectively. To improve the sampling, we used both the standard dithering and micro-dithering\footnote{\url{https://www.cfht.hawaii.edu/Instruments/Imaging/WIRCam/specsinformation.html}} (with a 2$\times$2 micro-stepping pattern) for our observations.

\subsection{Photometric Data Reduction}
We process the data using a custom data reduction pipeline named {\tt PyPhot}\footnote{\url{https://github.com/PyPhot/PyPhot}}.  {\tt PyPhot} includes the standard imaging data reduction processes including bias subtraction, flat fielding, and sky background subtraction.  

For the LBC images, the master bias and flat are generated by performing the sigma-clipped median on a series of bias and sky flats, respectively.  For $i$- and $z$-bands, we further correct fringing by subtracting off a master fringe frame constructed from our science exposures.  The sky background is estimated using {\tt SourceExtractor} \citep{bertin} after masking out bright stars. The cosmic rays are masked using the Laplacian edge detection algorithm \citep{vandokkum01}. 

For WIRCam images, we start with the pre-processed individual image data delivered by CFHT (dark subtracted, flat field corrected, and with preliminary background subtraction).  A detailed description of the detrending of these images can be found on the CFHT WIRCam image detrending webpage \footnote{\url{https://www.cfht.hawaii.edu/Instruments/Imaging/WIRCam/IiwiVersion2Doc.html}}. These detrended images are then processed with {\tt PyPhot} for further background subtraction, bad pixel masking, and cosmic ray rejection as was done for the LBC images.  

For each of these data sets, the final mosaic is produced using {\tt SCAMP} \citep{bertin06} and {\tt Swarp} \citep{bertin02}.  Additionally, a mask is created in order to remove saturation spikes and bright foreground stars in the images.  The effective clean area of the coadded and masked LBC mosaic is 0.153 square degrees and that of the WIRCam mosaic is 0.181 square degrees.  Additionally, the pixel scale of the final mosaics are $0.224 ''/$pixel and $0.153 ''/$pixel for the LBC and WIRCam images, respectively.

\subsection{Photometric Calibration and Catalog Creation}
We perform object detection on each mosaic with {\tt SourceExtractor} by setting {\tt DETECT\_THRESH=1.5} and {\tt DETECT\_MINAREA=4}. To calibrate the photometric measurements, the individual exposures for LBC and WIRCam are calibrated to the Pan-STARRS \citep{chambers} and 2MASS \citep{Skrutskie} infrared photometric catalogs, respectively. We only use sources that have been detected in all five Pan-STARRS bands.  We further require that the difference between the {\tt Kron} magnitude and {\tt PSF} magnitude in the Pan-STARSS catalog are smaller than 0.3 magnitudes in all five bands and have a signal-to-noise ratio greater than 10 with no FLAGS from {\tt SourceExtractor} in our data.  Finally, we restrict these points to be in the magnitude range of $18<{\rm mag} <20$ in Pan-STARRS and $15.5<{\rm mag}<17.5$ in 2MASS to ensure the strongest correlation in magnitudes. This results in $\sim250$ sources for the Pan-STARRS calibration and $\sim150$ sources for the 2MASS calibration. 

Using these bright point sources, we calibrate the zero-points for each filter using color terms derived from standard stars.  To check the reliability of our calibrations, we compare our magnitudes against the reference magnitudes obtained from the Pan-STARRS and 2MASS photometric catalogs after applying color corrections for the bright stars.  We find that the standard deviations of the differences between our magnitudes and the reference magnitudes is $0.05,\,0.06,\,0.05,\,0.15,$ and $0.09$ for the $r$-,$i$-,$z$-,$Y$-, and $J$-bands, respectively.  This photometric accuracy is highly adequate for high-redshift Lyman break selections. 

We then merge the catalogs of all sources detected in the five bands by assuming that objects with distance greater than $1\farcs0$ are unique sources as the seeing for the LBT observations is around $1\farcs0$ arcsecond. Finally, we perform forced aperture photometry for all unique objects with {\tt Photutils} \citep{photutils}. The exposure times and magnitude limits with a $2\farcs0$ diameter aperture of the fully calibrated images are listed in Table \ref{photometry}.

\begin{deluxetable}{cccc}[t]
\centering
\tablecaption{\textbf{Summary of observational information used in this study.}}
\tablehead{\colhead{Filter} & \colhead{Central Wavelength} & \colhead{Exposure Time} & \colhead{3$\sigma$ Depth} }
\startdata
     & [\AA]      & [hr]          & [mag]      \\
\hline
$r$    & 6200       & 3.7           & 26.55           \\
$i$    & 7670       & 1.5           & 26.38           \\
$z$    & 9608       & 2.1           & 25.79           \\
\hline
$Y$    & 10240      & 6.0           & 25.65           \\
$J$    & 12518      & 6.9           & 25.26
\enddata
\end{deluxetable}\label{photometry}

\section{Lyman Break Galaxy Candidate Selection} \label{sec:selection}
\begin{figure}
    \centering
    \includegraphics[width=\linewidth]{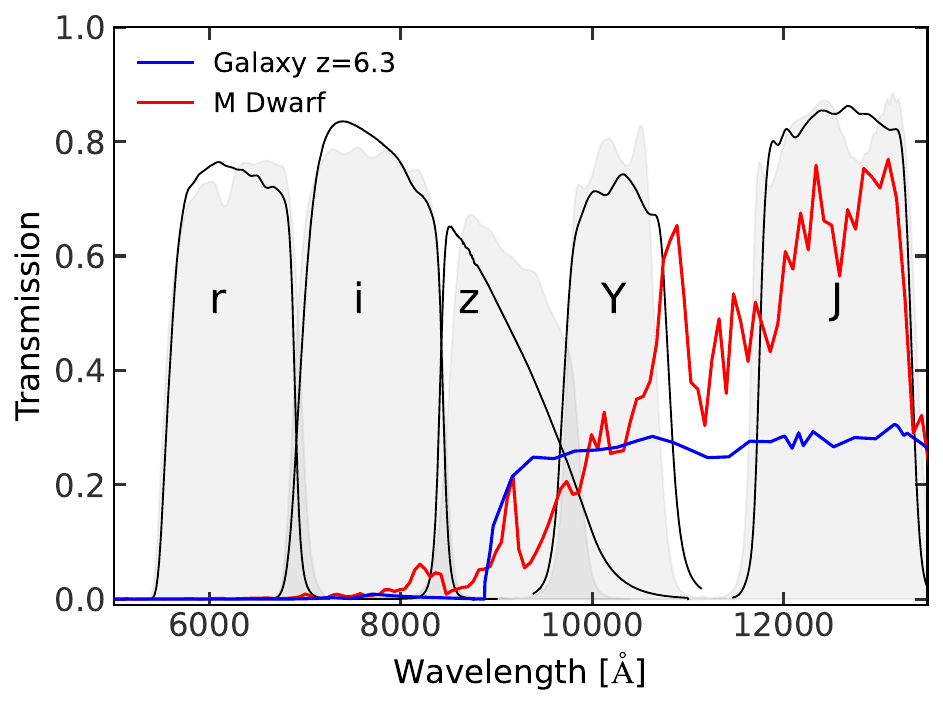}
    \caption{In black, transmission curves of the LBT/LBC $r$-, $i$-, and $z$-filters and of the CFHT/WIRCAM $Y$- and $J$-filters.  A template of a young star-forming galaxy redshifted to $z=6.3$ with IGM absorption taken into account \citep{inoue} is shown in blue with ($i$-$z$, $z$-$Y$, $z$-$j$) = ($3.85, 0.77, 0.82$).  An M-type dwarf stellar template is shown in red \citep{allard} with ($i$-$z$, $z$-$Y$, $z$-$j$) = ($2.37, 1.50, 2.67$). The grey, filled curves show the filter curves of the corresponding COSMOS data described in section 4.}
    \label{fig:temps}
\end{figure}

The Lyman break technique is an effective way to search for star-forming high-redshift galaxies due to the drastic decrease in flux observed at wavelengths blueward of the Ly$\alpha$ line ($\lambda_{\rm{Ly\alpha}}=1216$ \AA).  This drop in flux, due to the increasing neutral hydrogen fraction of the intergalactic medium (IGM) before the end of cosmic reionization causing photons with energies higher than Ly$\alpha$ to be absorbed, is known as a Gunn-Peterson trough \citep{gunn}.  At redshifts $z\geq 5$, this break in the spectra, usually seen in the ultraviolet (UV), is shifted into the near infrared (IR).  Specifically, at the redshift of J0100, Fig.~\ref{fig:temps} shows the Lyman break of a galaxy template shown in blue at a wavelength of $\lambda_{\rm{obs}}=8877$\AA\ which falls in the SDSS $z$ filter on the LBT. 

We use the Lyman break technique to select high-redshift galaxy candidates because it provides an effective selection criterion on photometric color rather than relying on many hours of spectroscopic observations.  However, as shown in Fig.~\ref{fig:temps} in red, this can be contaminated significantly by late-type M, L, and T dwarfs that have such red colors that they can appear to be dropouts.  Below, we explain the color criteria used to select these dropouts along with the color requirements and visual inspection procedure used to remove stellar and other contaminants.  

\subsection{Color-Color Diagram}

The following criteria are used as a preliminary selection of LBG candidates at $z\sim6$:
\begin{equation}
    z_{\rm{APER}}\leq 25.23 \hspace{0.2cm} {\rm and} \hspace{0.2cm} \left(S/N\right)_z > 5
\end{equation}
\begin{equation}
    \left(S/N\right)_r < 2
\end{equation}
\begin{equation}
    i -z > 1.5
\end{equation}

\begin{figure*}[t]
    \centering
\includegraphics[width=\linewidth]{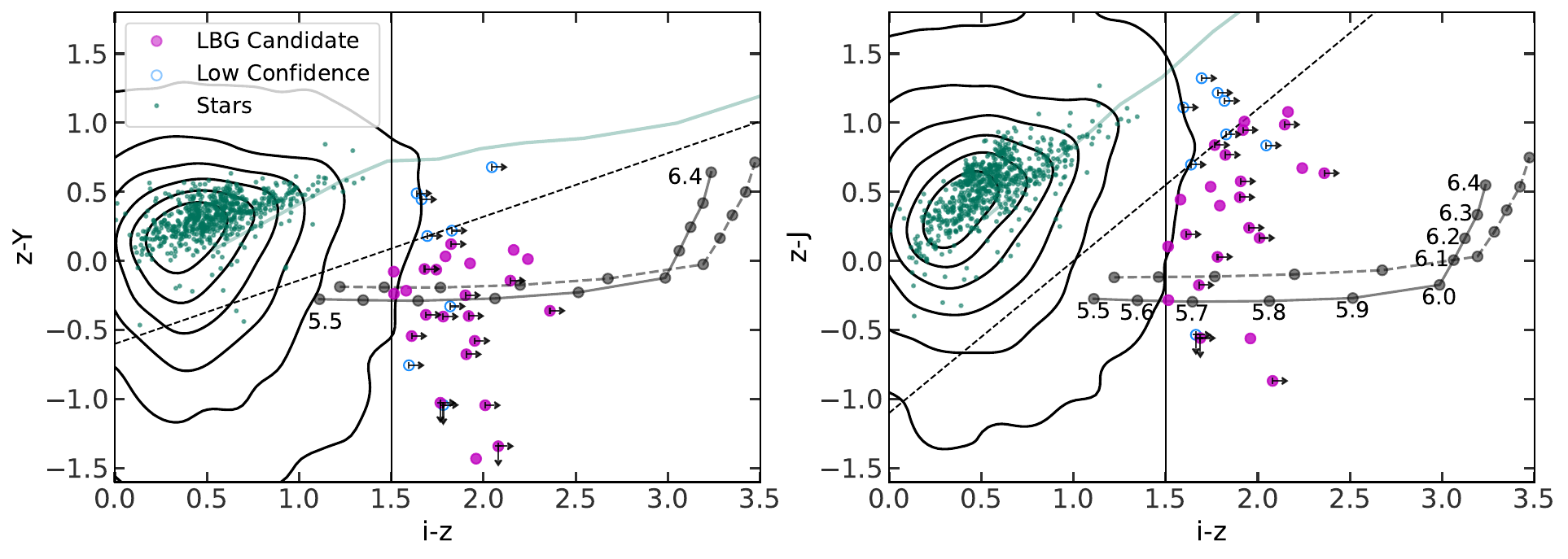}
    \caption{Color-color diagrams with LBG candidates (filled magenta points), low-confidence candidates (open blue circles), and stellar sources (green points) in the field.  These show the $z$-$Y$ (left) and $z$-$J$ (right) colors versus the $i$-$z$ color.  The vertical line shows the color cut at  $i$-$z<1.5$ while the diagonal dotted line shows the cut to remove stellar contaminants.  The grey tracks in each plot show the theoretical colors of a young star-forming galaxy at redshifts ranging from $z = 5.5$ to $z=6.4$ at redshift intervals of $\Delta z=0.1$.  The black contours show the region where all sources detected in the J0100 field reside.  The green track shows theoretical colors of solar metallicity MLT dwarfs calculated using the Sonora Model grid \citep{Marley}. Arrows on the pink and blue points show lower(upper) limits on the $i$-$z$($z$-$Y/J$) colors due to nondetections in the $i$, $Y$, or $J$ bands. Note, many of these candidates have true $i$-$z$ colors that are more red than portrayed.}
    \label{fig:colors}
\end{figure*}

The first two criteria, Eq. (1), require that the source must be detected in the \textit{z}-band to 5$\sigma$ and have a signal-to-noise ratio of greater than 5 in this band.  We use the $2\farcs0$ diameter aperture photometry for this cut, however, after further inspection, using the MAG\_AUTO magnitudes as other searches have done \citep[e.g.,][]{balmaverde, morselli}, does not change the results of this selection significantly.  The third criterion, Eq. (2), constrains the selection to sources not detected significantly in the $r$-band as the IGM absorption should be fully saturated at these wavelengths.  

Finally, as can be seen by the vertical line in Fig.~\ref{fig:colors}, the color selection in Eq. (3) favors sources whose Lyman Break falls at redshifts above $z\sim5.6$.   It has been shown that this is an efficient color cut for selecting starburst galaxies near redshift six \citep{stanway, bowler}.  For context, Fig.~\ref{fig:colors} also shows color tracks of a star-forming galaxy template simulated by the Flexible Stellar Population Synthesis  \citep{conroy09, conroy10} and retrieved from the EAZY code \citep{brammer, eazy}.  These tracks are shown for redshifts between 5.5 to 6.4 as grey lines.  

It is evident that we expect a non-detection in the \textit{i}-band filter.  To account for this non-detection, for any source that has less than a 2$\sigma$ detection in this filter, we use the 2$\sigma$ magnitude limit of $m_i=26.82$ as an upper limit to the corresponding sources in the \textit{i}-band.  This upper limit is used in calculating the $i-z$ colors and results in a lower limit in the $i-z$ colors (i.e. the true color is more red).  Even with the upper limit on the $i$-band magnitude, combining the first and fourth criteria means that it will still be eligible for color selection.  One caveat to consider is that extremely deep $i$-band imaging is needed to rule out $z=5.7$ galaxies from the overdensity.  We cannot definitively conclude that the overdensity is at the quasar's redshift or a redshift of $5.7$, therefore, we aim to attain spectroscopic followup of these galaxy candidates in the future.

After these four selection criteria are applied, 149 sources remain.  Many of which are low-redshift contaminants or spurious artifacts (e.g., bright star halos, saturation spikes, and cosmic rays). We also acknowledge that this color selection can result in a large redshift range ($5.7$-$6.5$) possibly probing galaxies that are not actually part of the same structure.  This is due to the use of broad-band photometric filters.  However, \citet{overzier09} showed that galaxy protoclusters can span up to 100 cMpc which corresponds to a window of roughly $\Delta z\sim 0.3$ centered on the quasar's redshift.  Thus, it is still possible that galaxies with a slight redshift offset are still within the same overdense structure. Additionally, a positive angular correlation of LBGs, even in a wide redshift window, can be used as evidence for being part of the same structure as is described in Section~\ref{sec:overdensity}.

\subsection{Removal of Contaminants}
\label{sec:subselect}
As mentioned above, low-mass stars, brown dwarfs, and Balmer Break galaxies at $z\sim1.5$ can contaminate this selection due to their extremely red colors \citep{bowler}.  Fig.~\ref{fig:temps} shows the comparison between a galaxy template at a redshift of $z=6.3$ in blue and an M-dwarf star in red.  It can be seen that this cool star has minimal flux in the $r$- and $i$-filters with some flux in the $z$-band similar to $i$-dropouts.  

\begin{figure*}[t]
\centering
\includegraphics[width=\linewidth]{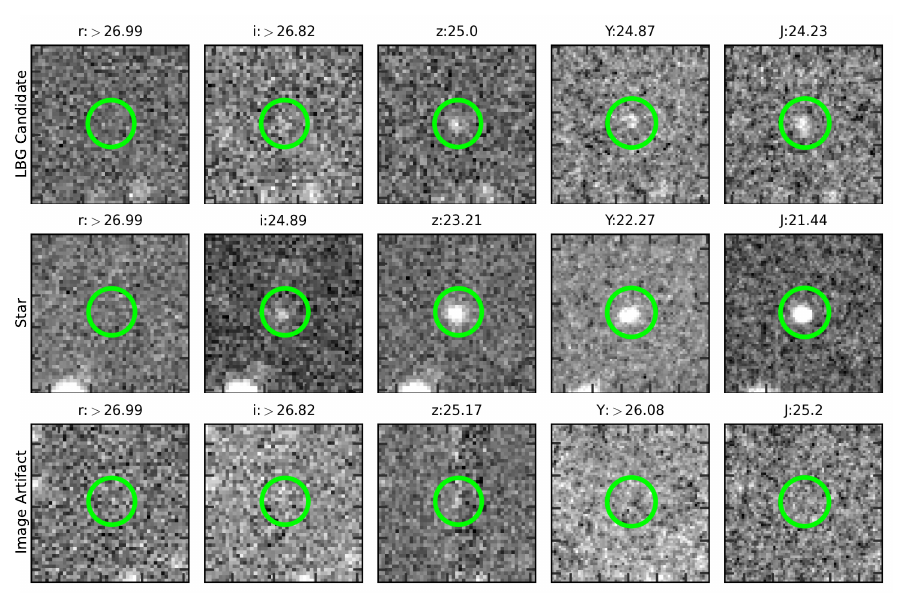}
    \caption{Examples of the visual inspection process using $10''\times10''$ cutouts of each source in the $r$, $i$, $z$, $Y$, and $J$ bands. Each cutout is independently normalized to a stretched Z-scale interval.  The top row displays a promising LBG candidate that passes visual inspection, the middle row shows a stellar source, and the bottom row shows an image artifact that is not a true source.  }
    \label{vis}
\end{figure*}

With the additional data given by the $Y$- and $J$-bands from the CFHT, it is possible to remove most of these targets as they should appear much redder in the $z-Y$ and $z-J$ colors.  Fig.~\ref{fig:colors} shows stars in the J0100 observations as green dots.  These are selected with the {\tt SourceExtractor} parameter CLASS\_STAR$>$0.98 and a magnitude limit of $z_{\mathrm{AUTO}}<23$.  As expected, these points generally populate a different color space than the color tracks of galaxies. 

This separation can also be seen as the green track in Fig.~\ref{fig:colors} which shows the typical colors of MLT dwarfs based on the Sonora model atmosphere grid \citep{Marley}. The diagonal lines in Fig.~\ref{fig:colors} show the relative color cut utilized to remove stellar contamination.  These cuts are optimized with the COSMOS data set \citep{cosmos} which provides a much larger set of stars determined with a higher confidence due to the many filters included in the COSMOS survey.  To be conservative while determining these cuts, we prioritize the purity of the sample over the completeness considering a high contamination fraction could result in a false overdensity signal.  To parameterize this cut, we first fit a line to the COSMOS stellar sources in the color-color plane.  Then the y-intercept is shifted far enough below the stellar locus to remove the majority of COSMOS stellar sources from the selection.  With this in mind, these diagonal cuts on the $z-Y$ and $z-J$ colors removed $99.8\%$ of the sources that are flagged as stars in the COSMOS catalog.  The following diagonal cuts are applied to the remaining 149 candidates from Section 3.1:
\begin{equation}
    z-J < 1.10(i-z)-1.1
\end{equation}
\begin{equation}
    z-Y < 0.46(i-z) - 0.6
\end{equation}

After eliminating stellar-like objects with the intersection of these two cuts (i.e. \textit{both} $z$-$Y$ and $z$-$J$ below the cut), 68 candidates are ready for visual inspection.  Due to the depth of the $i$-band filter, there are some sources that do not meet this color criteria but could be pushed out of the disallowed region due to the lower limit on the $i$-$z$ color.  Thus, we provide a low-confidence sample of possible LBG candidates that have a lower limit in the $i$-$z$ color, and have $z$-$Y$ \textit{or} $z$-$Y$ below the cut, and $J>23.5$.  This results in an additional 9 low-confidence sources to be visually inspected.

Visual inspection is required due to spurious sources and other image artifacts such as bright star halos and saturation spikes being incorrectly identified as sources by {\tt SourceExtractor}. An example of the result of this visual inspection can be seen in Fig.~\ref{vis} where the top row shows a valid LBG candidate, the middle row shows a stellar source (that was removed through the cuts of the $z$-$Y/J$ colors), and the bottom row shows a saturation spike erroneously extracted by {\tt SourceExtractor}.  

After visual inspection to remove extended sources and other defect/noise sources, 23 galaxies remain with an additional 8 low-confidence sources.  Thus, we discover 23 LBG candidates around J0100 in an area of $\sim 500$ square arcminutes, all of which can be seen in Fig.~\ref{fig:field}.  Their coordinates and magnitudes are tabulated in Appendix~\ref{tab:append} and the cutouts in each filter are shown in Appendix~\ref{appendB} at the end of this paper.  The additional low-confidence sources can be found at the bottom of the same table.  These low-confidence sources are not included in the calculations of overdensity or clustering that follow.

\begin{figure*}[t]
    \centering
\includegraphics[width=0.8\linewidth]{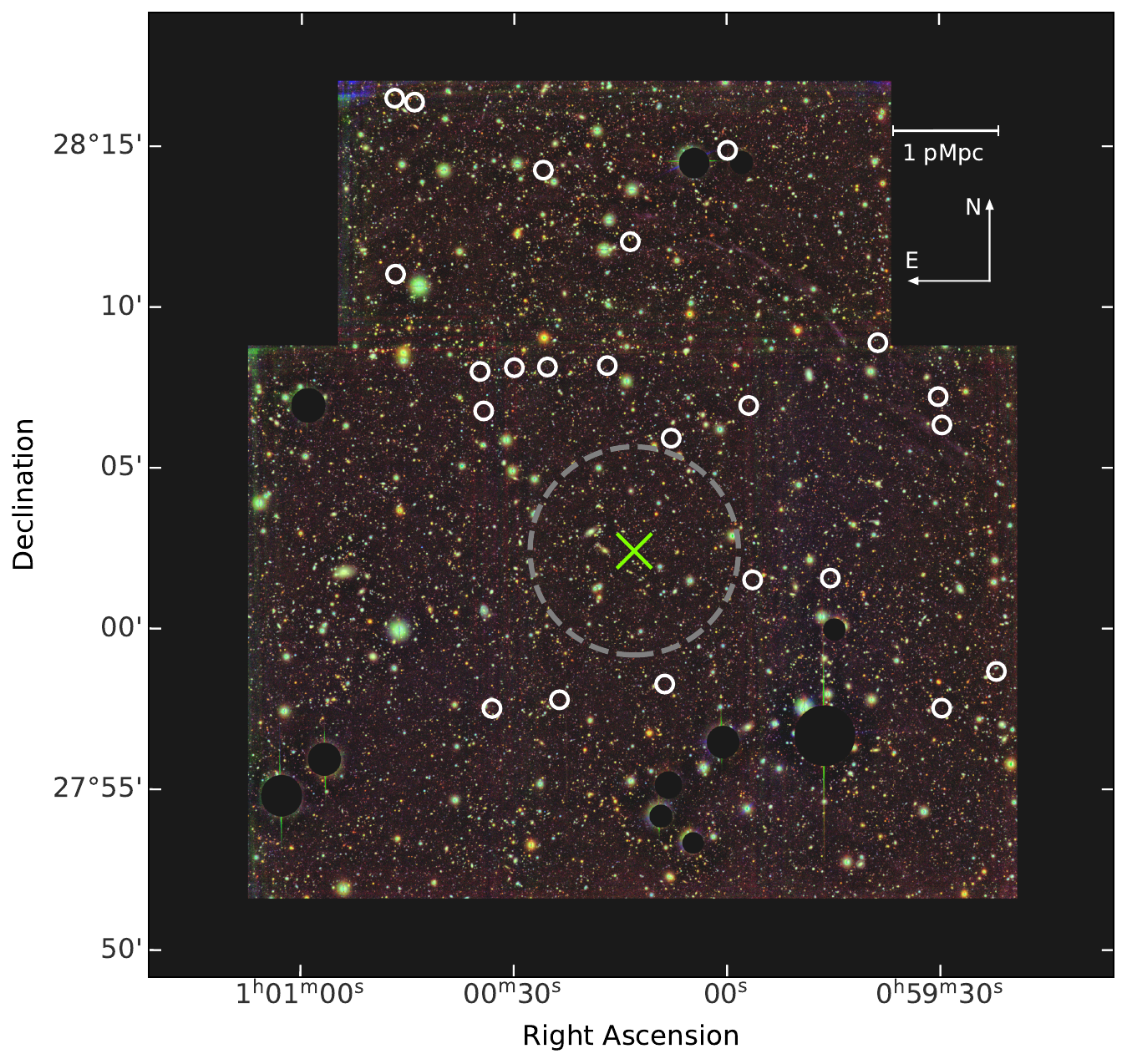}
    \caption{Spatial distribution of candidate LBGs as white circles superimposed on a composite color image of the quasar field.  The quasar location is marked with a green cross.  The field spans a $\sim23'\times25'$ area with the majority of the candidate LBGs located in the northwest region of the environment.  The grey dashed circle shows a region with a radius of 9 cMpc (1.2 pMpc) in which no LBG candidates are found.}
    \label{fig:field}
\end{figure*}

Of the 23 main candidates and 8 low-confidence candidates, none of them overlap with the \citet{kashino} [\ion{O}{3}] emitters in the same field.  In \citet{matthee}, in which these [\ion{O}{3}] emitters are characterized, the representative UV magnitude is ${M_{UV}=-19.6\pm0.1}$ for the full sample and ${M_{UV} = -19.5\pm 0.1}$ for the [\ion{O}{3}] emitters at $z>6.25$.  At the assumed redshift, this corresponds to an apparent magnitude of $M_{UV}\sim 27.2$ which is below the detection thresholds for our study.  Therefore, we do not expect to detect these as LBG candidates from our data.  Furthermore, the LBG candidates found here fall outside of the smaller FoV of the JWST observations at larger spatial scales.
\subsection{Spectral Energy Distribution Fitting}
We calculate the photometric redshifts with two codes - LePhare\footnote{\url{https://www.cfht.hawaii.edu/~arnouts/LEPHARE/lephare.html}} \citep{arnouts, ilbert} and EAZY\footnote{\url{https://github.com/gbrammer/eazy-py}} \citep{brammer, eazy} - in order to compare the $\chi^2$ values between galaxy and stellar templates.  Both programs fit spectral templates to the observed $2\farcs0$ aperture fluxes of each source.
For EAZY, we use the 17 templates adopted by  \citet{cosmos} with a uniform redshift prior.  LePhare uses both 31 galaxy templates \citep{ilbert09} and 254 stellar SEDs \citep{pickles, chabrier}.  The redshifts have large uncertainty due to the broad filters and small number of filters used.  However, the stellar fit $\chi^2$ values are indeed all larger than those for the galaxy fits indicating that these are likely high redshift galaxies as opposed to low-mass stars within the Galaxy.

It is worth noting that \citet{kashino} found several overdensities of [\ion{O}{3}] emitters with slight redshift offsets from J0100.  It is necessary for spectroscopic follow-up to determine if these candidate LBGs reside in the environment of the quasar or in foreground overdensities.  \citet{mignoli} has done this type of spectroscopic follow-up with LBGs found around another $z=6.3$ quasar J1030+0524 in \citet{balmaverde} using similar selection criteria as this paper.  Spectra were taken of 12 of the candidate LBGs confirming 9 high-redshift galaxies with 3 undetermined redshifts due to low spectral resolution and the lack of emission lines.  Thus, this selection has been proven to be robust and can be confidently used to find high redshift LBGs.  Overall, all candidate galaxies are unlikely to be low-redshift contaminants based on the stellar $\chi^2$ values. 


\section{Blank Field Comparison with COSMOS}
\label{expect}

The number of selected LBG candidates in the 0.153-square degree LBT field around J0100 is 23.  In order to put this number into context, we calculate the number of dropout galaxies in a large blank field using the same selection techniques.  We take advantage of the COSMOS field from \citet{cosmos} which provides a large amount of photometric data.  We apply the \texttt{COMBINED} flag ensuring that the sources in this area are covered by UltraVISTA, Suprime-Cam, and Hyper Suprime-Cam along with being free of edges and bright stars.  This results in a total area of 1.27 square degrees which is large enough to represent a field governed  by cosmic variance rather than any single overdensity at redshift 6.  Not only does this choice benefit from the large area of the COSMOS field, but it is also covered by a large selection of filters.  

\subsection{Blank Field LBG Surface Density}  
In order to make a comparable selection of $i$-dropout galaxies in the COSMOS field, it is necessary to choose filters that are the most similar to the LBT and CFHT filters used in this paper with depths that are similar to or deeper than our data.  We choose the $r$-, $i$-, and $z$-filters from Subaru's Suprime-Cam (SC) and the $Y$- and $J$-filters from VIRCAM on the VISTA telescope.  These filters and their respective $3\sigma$ depths are shown in Fig.~\ref{fig:temps} and Table~\ref{cosmos}.  The SC $z$-band has a similar depth to our data and therefore is ready to be used in the analysis.  However, the SC $r$- and $i$- bands are about a half magnitude deeper than this study.  Thus, it is necessary to degrade the COSMOS data in order to match the data quality of this paper.  
\begin{deluxetable}{cccc}[t]
\centering
\tablecaption{\textbf{Selected COSMOS Catalog Information} The source telescope, central wavelength, and $3\sigma$ depths for each filter of the COSMOS data used as a comparison to the LBT data.  }
\tablehead{\colhead{Filter} & \colhead{Source} & \colhead{Central Wavelength} & \colhead{3$\sigma$ Depth}}
\startdata
   ...    &   ...     & [\AA]      & [mag]            \\ \hline
$r$    & Subaru/SC     & 6305       & 27.1           \\
$i$    & Subaru/SC     & 7693       & 26.7           \\
$z$    & Subaru/SC     & 8978       & 25.7           \\\hline
$Y$    & VISTA/VIRCAM     & 10216      & 25.3           \\
$J$    & VISTA/VIRCAM     & 12525      & 25.9           
\enddata
\tablecomments{$Y$ and $J$ depths are for the Deep observations (not UltraDeep) stripes in the COSMOS field.}
\end{deluxetable}\label{cosmos}

\begin{figure}[t]
    \centering
\includegraphics[width=\linewidth]{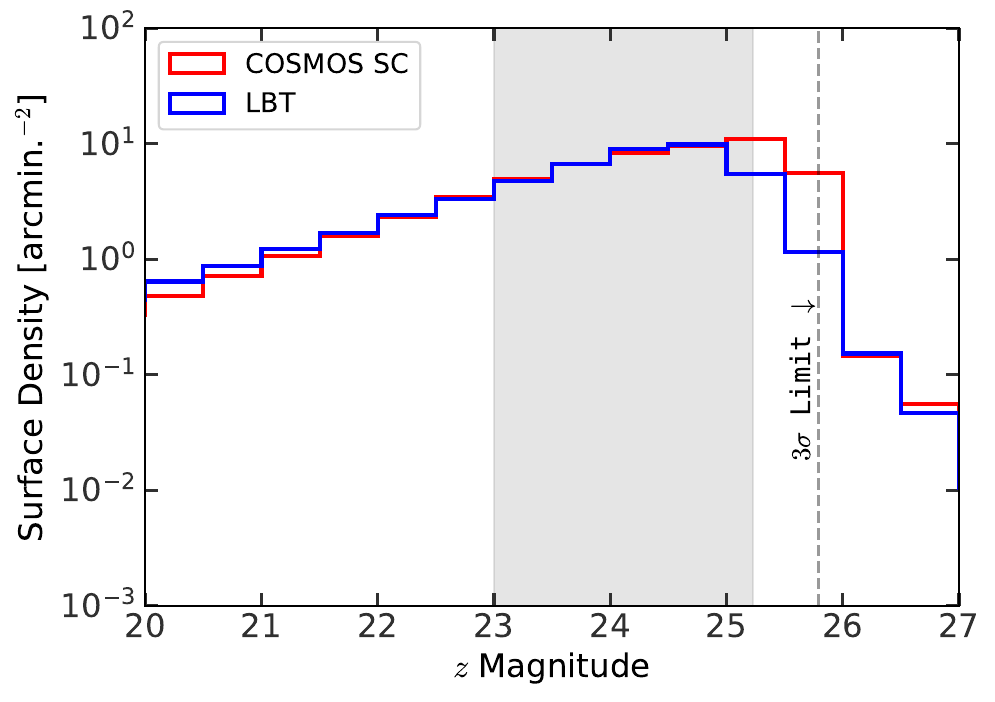}
    \caption{Surface density per square arcminute of the \textit{total number }of sources detected in the $z$-band in the LBT field (blue) and the COSMOS field (red).  The shaded grey region shows the magnitude range of interest within which there is excellent agreement in completeness between the two fields showing that it is appropriate to use the COSMOS field as a comparison for the overdensity calculation. }
    \label{fig:zmags}
\end{figure}

To fulfill this requirement, we match the background flux limits of the COSMOS data with that of the LBT data in each filter.  We convert the SC $3\sigma$ magnitude limit to a flux limit for each filter and subtract this value in quadrature from the original flux errors of the SC sources.  Next we add in quadrature the $3\sigma$ flux limit of the LBT filters to the errors.  
With this done, the distributions of errors in the relevant flux range ($0.1-1\mu$Jy) are the same between the COSMOS and LBT data. To degrade the flux values, we add a Gaussian distributed noise term to the fluxes with mean zero and standard deviation of $\sigma = \sqrt{\sigma_{LBT,sky}^2-\sigma_{COS,sky}^2}$, where $\sigma_{*,sky}$ are the $3\sigma$ flux depths of the images.  From here, we convert back to magnitudes in order to proceed with the same candidate selection as described in section~\ref{sec:selection}.  Fig.~\ref{fig:zmags} shows the distribution of $z$ magnitudes, the detection filter, for each field.  After following the same selection analysis -- which involves the color selections for redshift ($i$-$z$) and contaminants ($z$-$Y$/$z$-$J$), visual inspection, and photometric redshift calculations -- we detect 34 $i$-dropout galaxy candidates in the whole COSMOS field.  That is, we expect 0.007 LBGs per square arcminute. 

\subsection{Completeness and Purity}
One benefit of using the COSMOS field is the ability to check the completeness and contamination fraction of our selection using the published photometric redshifts in the catalog.  COSMOS uses over 20 filters ranging from the UV to NIR to fit to galaxy templates using EAZY and Lephare.  This abundance of filters allows for a more accurate photometric redshift to be determined. We use the photometric redshifts from the full COSMOS dataset as a proxy of the true spectroscopic redshifts, and compare the results to the selection using data degraded to match the LBT/WIRCam dataset.

Using the full COSMOS dataset, we find 48 galaxies with photometric $z>5.5$ to a flux limit of $z_{AB} =25.23$. 
Of these 48 high-redshift galaxies, we select 17 objects using degraded data following the same selection criteria.  Thus, the completeness of our selection technique is 35.4\% meaning there could be up to 60 more LBGs in the environment of the quasar at this flux limit.  

Promisingly, of the 34 LBG candidates selected by the degraded COSMOS data using only five filters, 50\% of them (17) are designated true high-redshift galaxies with the COSMOS redshifts, while others are low-redshift interlopers due to the degraded data quality.  This 50\% contamination rate indicates that at least 13 candidates in the J0100 field are true high-redshift galaxies in the environment of the quasar.

\section{Overdensity and Structure of LBGs around J0100} \label{sec:overdensity}
\subsection{Overdensity Measurement}
The overdensity of galaxies in a portion of the sky is determined by the equation 
\begin{equation}
    \delta =  \frac{n}{\bar{n}}-1.
\end{equation} 
In this equation, $\bar{n}$ is the average number of galaxies in a blank field where no overdensity is expected, and $n$ is the number of galaxies actually detected in the area of the quasar.  

\begin{figure}[t]
    \centering
    \includegraphics[width=\linewidth]{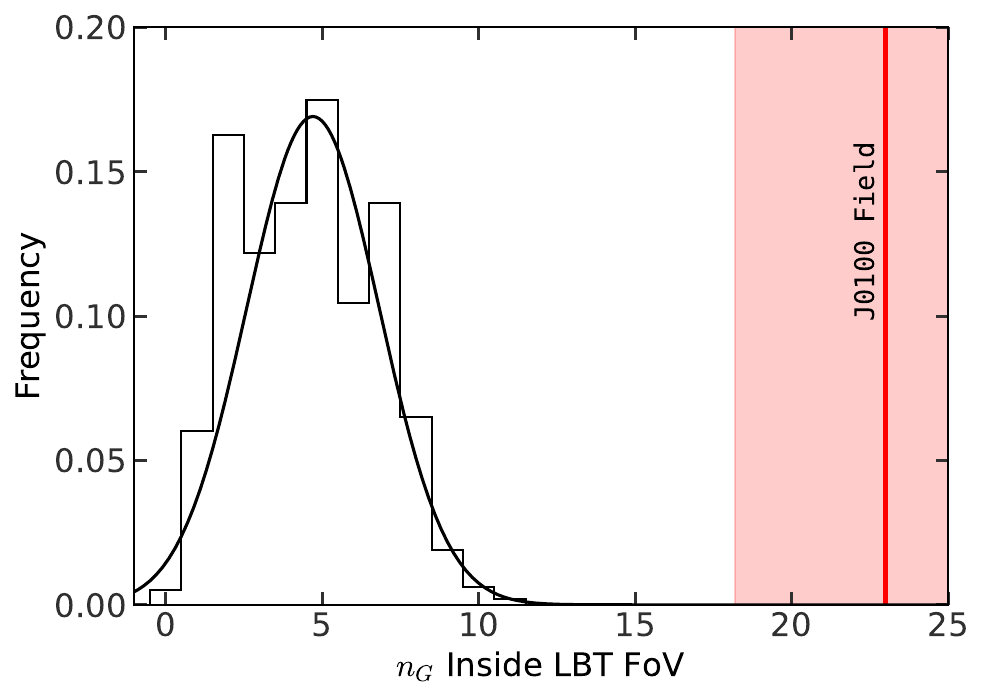}
    \caption{Frequency histogram of the number of LBGs counted in an LBT FoV-sized box after 10,000 Poisson pointings within the COSMOS field.  The thick black line is a Gaussian fit to the data with $\mu=4.6$ and $\sigma=2.2$.  The red vertical line shows the number of LBG candidates counted in the J0100 field while the red shaded region shows the Poisson error on this count.}
    \label{fig:poiss}
\end{figure}

To calculate the number of candidates from the COSMOS field expected in a field size of the LBT FoV, we randomly point a box with the same dimensions of the LBT FoV at the COSMOS field 10,000 times.  For each pointing, the number of the 34 selected LBG candidates from the COSMOS field within the area is recorded.  We fit a Gaussian function to the distribution of these counts and recover a mean galaxy count of 4.6 with a standard deviation of 2.2. The distribution of these pointings is shown in Fig.~\ref{fig:poiss} where the red vertical line shows the number of candidate galaxies in the J0100 field.  

It is evident that there is a significant overdensity in the field around J0100.  The contamination rate does not affect this calculation, because the expected number of LBGs, $\bar{n}$, is calculated from degraded COSMOS data with the same effect.  Specifically, with the expected counts, we calculate an overdensity of $\delta=(23/4.6)-1=4$ at 8.4$\sigma$ significance in the field of J0100. 

\subsection{Spatial Distribution and Angular Correlation}
As the photometric redshifts derived using the EAZY code with only 5 filters have large uncertainty, it is not reliable to map this protocluster in 3D space using our data.  Rather, the calculation of the 2D two-point angular auto-correlation function (ACF) can indicate clustering on an angular scale.  If the overdensity is due to a chance alignment of galaxies along the line of sight, one would expect to see no angular clustering above that of a blank field.  However, if the galaxies within the protocluster are truly associated with one another, one would expect them to be strongly clustered.   

Fig.~\ref{fig:field} shows the distribution of the LBG candidates in the field and highlights the need for large FoVs as there are no LBG candidates within 9 cMpc (diameter of $7.4'$) from the quasar.  Using single pointings from Hubble's Advanced Camera for Surveys ($\sim3.4'\times3.4'$) or JWST's NIRCam ($\sim2.2'\times5.1'$), one would: at best, not capture the full extent of the galaxy overdensity, and at worst, not detect it at all.  

Additionally, it is evident from Fig.~\ref{fig:field} that many of the galaxy candidates reside in the northwest portion of the imaging field.  To ensure that this is not due to sensitivity variations in the different chips on the detector or other sky variations, we calculate the number counts from the original catalog with quality cuts in each quadrant of the image.  The results show that while the total catalog and clean catalog show the same distribution of sources in each quadrant (roughly $25\%$ as expected), the candidate distribution does not.  Running a two-sample Z-test between the percent of sources in the northwest quadrant from the catalog and that of the candidates shows that there is only a $1.1\%$ chance that there would be this large of a fraction of candidates in this quadrant compared to the original distribution of sources in the image.  This indicates that the asymmetry of the candidate distribution is not likely due to the distribution of the original catalog. 

To evaluate this structure in a more quantitative way, we use the two-point angular  ACF, which calculates the likelihood of finding a galaxy within a goven angular distance of another galaxy compared to what would be expected from a randomly distributed population.  To measure this, we use the \citet{Landy} correlation function estimator which is used in many galaxy correlation studies \citep{ overzier06, lee,mclure, wangcorr}.  This takes the form of 
\begin{equation}
    \omega(\theta) = \frac{\widehat{DD} - 2\widehat{DR} + \widehat{RR}}{\widehat{RR}}.
\end{equation}
$\widehat{DD}$, $\widehat{DR}$, and $\widehat{RR}$ are the \textit{normalized} pair counts between real galaxies, real galaxies and random points, and random points residing within separations of $\theta+\Delta \theta$.  These are calculated with the raw number of galaxy-galaxy, galaxy-random, and random-random pairs ($DD$, $DR$, $RR$, respectively), the number of data sources ($n_d$), and the number of random points ($n_r$) as follows:
\begin{eqnarray}
    \widehat{DD} &= \frac{DD}{n_d(n_d-1)/2}\\
    \widehat{DR} &= \frac{DR}{n_d n_r}\\
    \widehat{RR} &= \frac{RR}{n_r(n_r-1)/2}.
\end{eqnarray}

\begin{figure}[t]
    \centering
    \includegraphics[width=\linewidth]{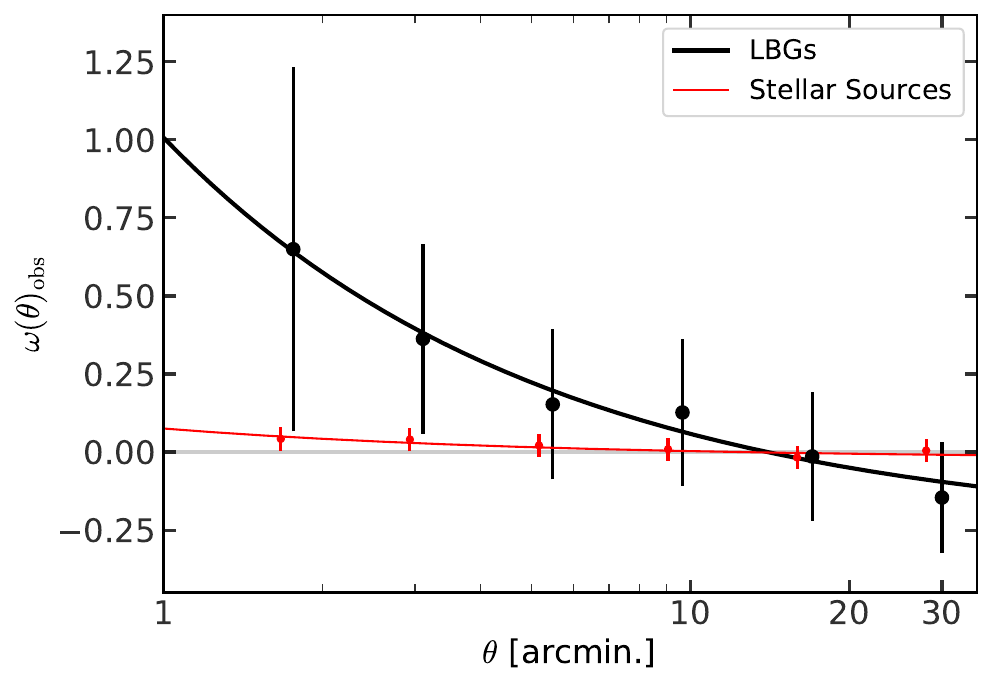}
    \caption{Angular auto-correlation function of the 23 LBG candidates in the J0100 field (black points) along with the same for the stellar contaminants shown as red points.  Solid lines represent the power law ($\beta=0.6$) fits to the data with $A_{\omega, \mathrm{LBG}} = 15$ and $A_{\omega, \mathrm{stellar}} = 1.1$.}
    \label{fig:angcorr}
\end{figure}

We use a random distribution made up of 10,000 mock sources that fit within the same geometry of the LBT field.  This takes into account the mask used to remove areas around saturated stars and other noisy regions used during the selection process.  We investigate the span of separations between $1'-30'$.  These angular separations correspond to $\sim2-75$ cMpc at the redshift of the quasar, thus probing the protocluster to its outermost regions \citep{overzier09}.  We use logarithmic binning of these separations in order to sufficiently sample the pairs at small separations while avoiding unnecessarily fine binning at large angular separations as seen in Fig.~\ref{fig:angcorr}.  

The ACFs for both the 23 LBG candidates and for stellar sources in the LBT field are shown in Fig.~\ref{fig:angcorr} in black and red, respectively.  Errors were determined using the Poisson estimator 
\begin{equation}
    \Delta \omega(\theta) = \frac{1+\omega(\theta)}{\mathrm{min}(N_{\mathrm{pair}}, N_{\mathrm{data}})^{1/2}}
\end{equation} 
as described in \citet{croom} and \citet{daangela}. The galaxy candidates show a positive clustering signal at separations of less than $10'$.  The stellar sources in the field, determined by CLASS\_STAR$\geq$ 0.98, show essentially no clustering signal as expected for a random distribution of stars in the FoV.  

We assume a power-law ACF in the form $\omega_{\mathrm{obs}}(\theta) = \omega_{\mathrm{true}}(\theta) - \mathrm{IC} = A_{\omega}\theta^{-\beta} -\mathrm{IC}$ in which the observed ACF signal is skewed downward due to the finite geometry of the field \citep{roche}.  The integral constraint, $\mathrm{IC}$, is used to fit for this underestimation of the clustering signal and is calculated using 
\begin{equation}
    \mathrm{IC} = \frac{\sum RR(\theta)\theta^{-\beta}}{\sum RR(\theta)}.
\end{equation}  
The resulting correction used is $\mathrm{IC} = 0.018$.  We fit the data using this formula and taking $\beta = 0.6$ as is used in many clustering analyses \citep{overzier06, lee}.  The resulting amplitude for the galaxy candidate distribution is $A_{\omega} = 15\pm 2\,\mathrm{arcsec.^{0.6}}$ and for the stellar distribution $A_{\omega} = 1.1\pm 0.2\,\mathrm{arcsec.^{0.6}}$.  \citet{overzier06}, \citet{lee}, and \citet{harikane} determined the clustering of $i$-dropout galaxies at $z\sim6$ in the GOODS fields and found clustering amplitudes of $2.71\pm2.05$, $1.12_{-0.25}^{+0.34}$, and $2.7\pm1.3$, respectively, for the bright galaxies in their samples.  These are an order of magnitude smaller than that of the galaxies in this study.   This strengthens the evidence that these candidate LBGs in the J0100 field come from the same overdense structure.  

\subsection{Results in Context}
Similar searches for LBGs on $\sim 10$ Mpc scales around $z\sim 6$ quasars have been conducted with similar results \citep[see for e.g.][]{utsumi,morselli, balmaverde, ota}.  Each study reports at least slightly overdense quasar fields revealing that searches on these large scales may provide a less biased view into the large-scale structure around quasars.  Additionally, those studies that applied extra photometric constrains (e.g. \citet{utsumi} uses the $z_R$ filter and \citet{balmaverde} uses the $Y-$ and $J$-bands) report slightly higher overdensities.  For example, \citet{morselli} looked at the J1030+0524 field only using the $r$-, $i$-, and $z$-bands and found an overdensity of $\delta = 2.0$; however, \citet{balmaverde} expanded upon this search with the $Y$- and $J$-bands and calculated an overdensity of $\delta = 2.4$ with more confidence of contamination removal.  J0100, having a black hole mass roughly 10 times that of J1030 \citep{derosa}, has a slightly higher overdensity using extremely similar selection techniques.  This could indicate a possible correlation with black hole mass, though, a much larger sample with spectroscopic confirmation is needed for this assertion.

The LBG candidates in each study are also distributed asymmetrically in the quasar fields.  The quasars are not necessarily found at the centers of these overdensities nor are they in the most dense regions.  Additionally, all of the observed quasar fields show a lack of LBG candidates in the direct vicinity of the quasar itself.  \citet{utsumi, morselli} and \citet{balmaverde} all report either no or very few LBGs within $1-3$ pMpc ($7-21$ cMpc at $z\sim6$).  Similarly, we find no LBG candidates within 1.2 pMpc  ($9$ cMpc) from J0100.  These values are on the order of the sizes of proximity zones around quasars: the point where the Ly$\alpha$ transmission drops below 10\% \citep{fan06}.  This indicates possible suppression of star formation in galaxies near the quasar due to UV radiation from the quasar heating the IGM causing faint galaxies to dominate at regions closest to the quasar. 

It is evident from these results, and those prior, that quasars are likely to live in overdense regions traced by LBGs.  While their properties are similar, their exact values (such as overdensity estimate, clustering signal, and proximity to the quasar) fill a range of parameter space that is still unconstrained.





\section{Summary and Conclusion} \label{sec:summary}
We utilize the wide-field ($\sim23'\times25'$) imaging in the $r$-, $i$-, and $z$-bands from the Large Binocular Camera on the Large Binocular Telescope along with complementary $Y$- and $J$-band imaging from the Wide-field Infrared Camera on the Canada-France-Hawaii Telescope to inspect the large-scale environment around the ultraluminous quasar, SDSS J010013.02+280225.8 (J0100).  This quasar is the most massive known quasar at $z\geq6$, making it an ideal region of space to search for large-scale structure traced by galaxies.  The spatial scales probed by this wide field of view correspond to a $\sim50\times 50$ cMpc$^2$ region, the expected extent of large protoclusters from simulations. 

We construct a catalog of sources in this field and utilize magnitude-, signal-to-noise ratio-, and color-based selection thresholds to identify $i$-dropout LBGs.  We find 23 high-confidence LBGs in the field while only 4.6 LBGs are expected in a region the size of the LBT FoV according to the COSMOS field matched to the data characteristics of our survey. This gives rise to a measured overdensity of $\delta=4$ at $8.4\sigma$ significance.  The candidate LBGs show clustering ($A_{\omega} = 15\pm 2\,\mathrm{arcsec.^{0.6}}$) an order of magnitude larger than foreground stellar sources furthering the evidence of large-scale structure around J0100.





Spectroscopic follow-up will be required on these candidate LBGs to determine their true redshifts and certify whether or not they are truly forming a large-scale structure around J0100. 

\begin{acknowledgments}
MP, FW snd XF acknowledge support from NSF grant AST-1908284 and AST-2308258. 

This research uses data obtained through the Telescope Access Program (TAP), which has been funded by the Strategic Priority Research Program “The Emergence of Cosmological Structures” (grant No. XDB09000000), National Astronomical Observatories, Chinese Academy of Sciences, and the Special Fund for Astronomy from the Ministry of Finance. 

Based on observations obtained with WIRCam, a joint project of CFHT, Taiwan, Korea, Canada, France, at the Canada-France-Hawaii Telescope (CFHT) which is operated from the summit of Maunakea by the National Research Council of Canada, the Institut National des Sciences de l'Univers of the Centre National de la Recherche Scientifique of France, and the University of Hawaii. The observations at the Canada-France-Hawaii Telescope were performed with care and respect from the summit of Maunakea which is a significant cultural and historic site.

The LBT is an international collaboration among institutions in the United States, Italy and Germany. LBT Corporation partners are: The University of Arizona on behalf of the Arizona university system; Istituto Nazionale di Astrofisica, Italy; LBT Beteiligungsgesellschaft, Germany, representing the Max-Planck Society, the Astrophysical Institute Potsdam, and Heidelberg University; The Ohio State University, and The Research Corporation, on behalf of The University of Notre Dame, University of Minnesota and University of Virginia.

\textit{Software:} NumPy \citep{harris2020array}, Pandas \citep{McKinney_2010, McKinney_2011}, Matplotlib  \citep{Hunter:2007}, Astropy \citep{2018AJ....156..123A, 2013A&A...558A..33A}, SciPy \citep{Virtanen_2020}, EAZY \citep{brammer}, LePHARE \citep{arnouts, ilbert}, Corrfunc \citep{corrfunc}, Healpy \citep{zonca}.

\end{acknowledgments}
\clearpage

\appendix
\restartappendixnumbering
\section{All LBG Candidate Properties}
\label{tab:append}
The following table reports the information of the 23 LBG candidates along with the information for the additional 8 low-confidence LBG candidates. 
\begin{deluxetable}{cccccccc}[h!]
\centering
\tablecaption{\textbf{LBG Candidate Photometry} The coordinates and 2\farcs0 aperture magnitudes of each of the 23 LBG candidates in our sample. Entries containing a `$>$' symbol represent nondetections of the source and are depicted as the $2\sigma$ upper limit for the corresponding filter.  The bottom 8 sources are the low-confidence candidates described in section~\ref{sec:subselect}.}
\tablehead{\colhead{ID} & \colhead{RA} & \colhead{DEC} & \colhead{$r$} & \colhead{$i$}& \colhead{$z$}& \colhead{$Y$}& \colhead{$J$} }
\startdata
14812                  & 00:59:22.05 & +27:58:40.42 & $>$26.99 & $>$26.82              & 24.81$\pm$0.17        & 25.85$\pm$0.57        & 24.64$\pm$0.26        \\
25390                  & 00:59:29.68 & +28:06:20.60 & $>$26.99 & $>$26.82              & 25.00$\pm$0.17        & 24.87$\pm$0.18        & 24.23$\pm$0.14        \\
13026                  & 00:59:29.77 & +27:57:32.16 & $>$26.99 & 25.76$\pm$0.20        & 24.18$\pm$0.08        & 24.39$\pm$0.10        & 23.74$\pm$0.09        \\
26465                  & 00:59:30.20 & +28:07:13.51 & $>$26.99 & 26.57$\pm$0.43        & 24.83$\pm$0.15        & 24.89$\pm$0.17        & 24.29$\pm$0.14        \\
28643                  & 00:59:38.69 & +28:08:54.29 & $>$26.99 & $>$26.82              & 25.14$\pm$0.21        & 25.20$\pm$0.19        & 25.31$\pm$0.30        \\
19190                  & 00:59:45.43 & +28:01:35.11 & $>$26.99 & $>$26.82              & 25.13$\pm$0.20        & 25.52$\pm$0.27        & $>$25.69              \\
19112                  & 00:59:56.38 & +28:01:31.74 & $>$26.99 & $>$26.82              & 24.87$\pm$0.17        & 25.45$\pm$0.25        & 24.63$\pm$0.17        \\
26129                  & 00:59:56.91 & +28:06:57.30 & $>$26.99 & $>$26.82              & 25.04$\pm$0.21        & 25.44$\pm$0.24        & 25.01$\pm$0.24        \\
36179                  & 00:59:59.89 & +28:14:53.37 & $>$26.99 & $>$26.82              & 24.67$\pm$0.13        & 24.82$\pm$0.17        & 23.69$\pm$0.10        \\
24874                  & 01:00:07.84 & +28:05:56.37 & $>$26.99 & $>$26.82              & 24.46$\pm$0.11        & 24.82$\pm$0.15        & 23.82$\pm$0.08        \\
14197                  & 01:00:08.76 & +27:58:17.56 & $>$26.99 & $>$26.82              & 24.74$\pm$0.13        & $>$26.08              & 25.61$\pm$0.42        \\
33271                  & 01:00:13.61 & +28:12:03.02 & $>$26.99 & 26.43$\pm$0.37        & 24.92$\pm$0.16        & 24.99$\pm$0.17        & 24.81$\pm$0.20        \\
27691                  & 01:00:16.86 & +28:08:11.82 & $>$26.99 & $>$26.82              & 25.05$\pm$0.19        & $>$26.08              & 24.22$\pm$0.11        \\
13448                  & 01:00:23.58 & +27:57:48.18 & $>$26.99 & 26.68$\pm$0.47        & 24.52$\pm$0.11        & 24.44$\pm$0.10        & 23.44$\pm$0.06        \\
27635                  & 01:00:25.29 & +28:08:09.21 & $>$26.99 & $>$26.82              & 24.91$\pm$0.17        & 25.59$\pm$0.26        & 24.34$\pm$0.12        \\
35552                  & 01:00:25.88 & +28:14:16.47 & $>$26.99 & $>$26.82              & 24.92$\pm$0.16        & 25.17$\pm$0.22        & 24.46$\pm$0.17        \\
27638                  & 01:00:29.96 & +28:08:07.76 & $>$26.99 & 25.62$\pm$0.21        & 24.10$\pm$0.09        & 24.34$\pm$0.08        & 24.39$\pm$0.13        \\
12974                  & 01:00:33.08 & +27:57:31.84 & $>$26.99 & 26.38$\pm$0.39        & 24.42$\pm$0.12        & 25.85$\pm$0.36        & 24.98$\pm$0.23        \\
25887                  & 01:00:34.28 & +28:06:47.10 & $>$26.99 & $>$26.82              & 24.90$\pm$0.18        & 25.30$\pm$0.20        & 23.95$\pm$0.09        \\
27460                  & 01:00:34.79 & +28:08:00.80 & $>$26.99 & 26.63$\pm$0.50        & 24.39$\pm$0.11        & 24.38$\pm$0.09        & 23.72$\pm$0.07        \\
37982                  & 01:00:44.07 & +28:16:23.24 & $>$26.99 & 26.45$\pm$0.49        & 24.52$\pm$0.13        & 24.54$\pm$0.17        & 23.51$\pm$0.10        \\
32277                  & 01:00:46.73 & +28:11:02.16 & $>$26.99 & $>$26.82              & 25.21$\pm$0.21        & 25.75$\pm$0.31        & 25.02$\pm$0.23        \\
38109                  & 01:00:46.88 & +28:16:30.54 & $>$26.99 & 26.15$\pm$0.42        & 24.35$\pm$0.12        & 24.32$\pm$0.15        & 23.95$\pm$0.16        \\ 
\hline
\multicolumn{8}{c}{Low-Confidence Sources}\\ 
\hline 
7456                   & 00:59:35.55 & +27:53:49.84 & $>$26.99 & $>$26.82              & 24.99$\pm$0.17        & 24.77$\pm$0.19        & 24.08$\pm$0.14  \\
32924                  & 01:00:09.07 & +28:11:42.97 & $>$26.99 & $>$26.82              & 25.22$\pm$0.21        & 25.98$\pm$0.41        & 24.11$\pm$0.11  \\
35445                  & 01:00:11.73 & +28:14:09.58 & $>$26.99 & $>$26.82              & 25.18$\pm$0.21        & 24.69$\pm$0.15        & 24.49$\pm$0.17  \\
34296                  & 01:00:16.06 & +28:13:03.00 & $>$26.99 & $>$26.82              & 25.03$\pm$0.18        & 25.33$\pm$0.24        & 23.84$\pm$0.09  \\
10255                  & 01:00:23.97 & +27:55:38.41 & $>$26.99 & $>$26.82              & 25.13$\pm$0.20        & 24.94$\pm$0.18        & 23.8$\pm$0.09   \\
33372                  & 01:00:37.53 & +28:12:09.23 & $>$26.99 & $>$26.82              & 25.15$\pm$0.20        & 24.71$\pm$0.12        & $>$25.69        \\
25884                  & 01:01:04.20 & +28:06:45.70 & $>$26.99 & $>$26.82              & 24.77$\pm$0.15        & 24.10$\pm$0.12        & 23.94$\pm$0.14  \\
23393                  & 01:01:05.40 & +28:04:47.87& $>$26.99  & $>$26.82              & 25.04$\pm$0.20        & $>$26.08             & 23.82$\pm$0.113 
\enddata
\end{deluxetable}\label{props}

\clearpage
\section{All LBG Candidate Cutouts}
\label{appendB}
\begin{figure}[ht]
    \centering
    \includegraphics[width=0.85\linewidth]{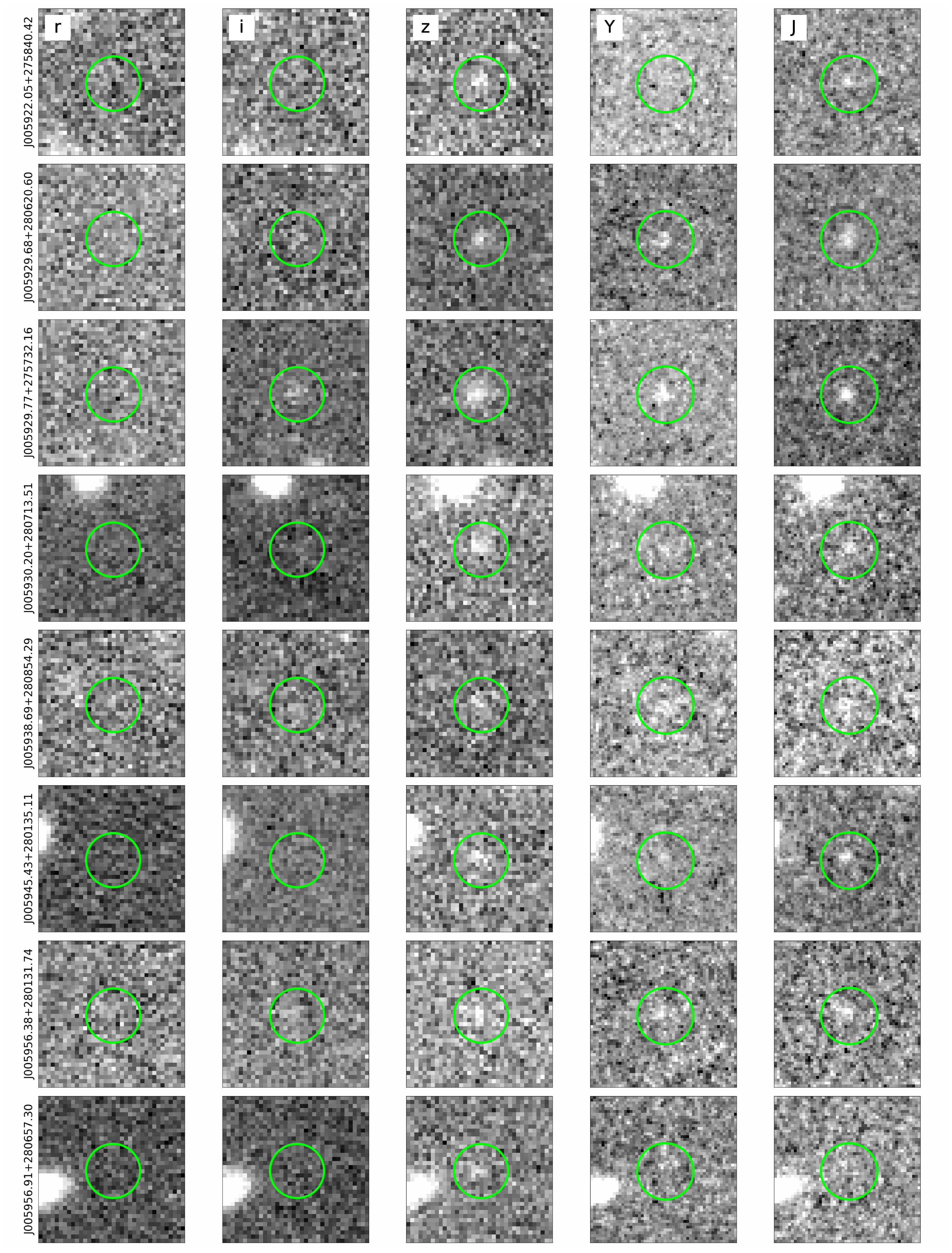}
    \caption{Postage stamp cutouts of each high-confidence candidate LBG listed in Appendix~\ref{tab:append}.  These show a 8$\times$8 square arcsecond cutout centered on each target.  The green circles are to aid the eye and have diameters of 3 arcseconds.}
    \label{fig:cuts}
\end{figure}

\begin{figure}[ht]
    \centering
    \includegraphics[width=0.85\linewidth]{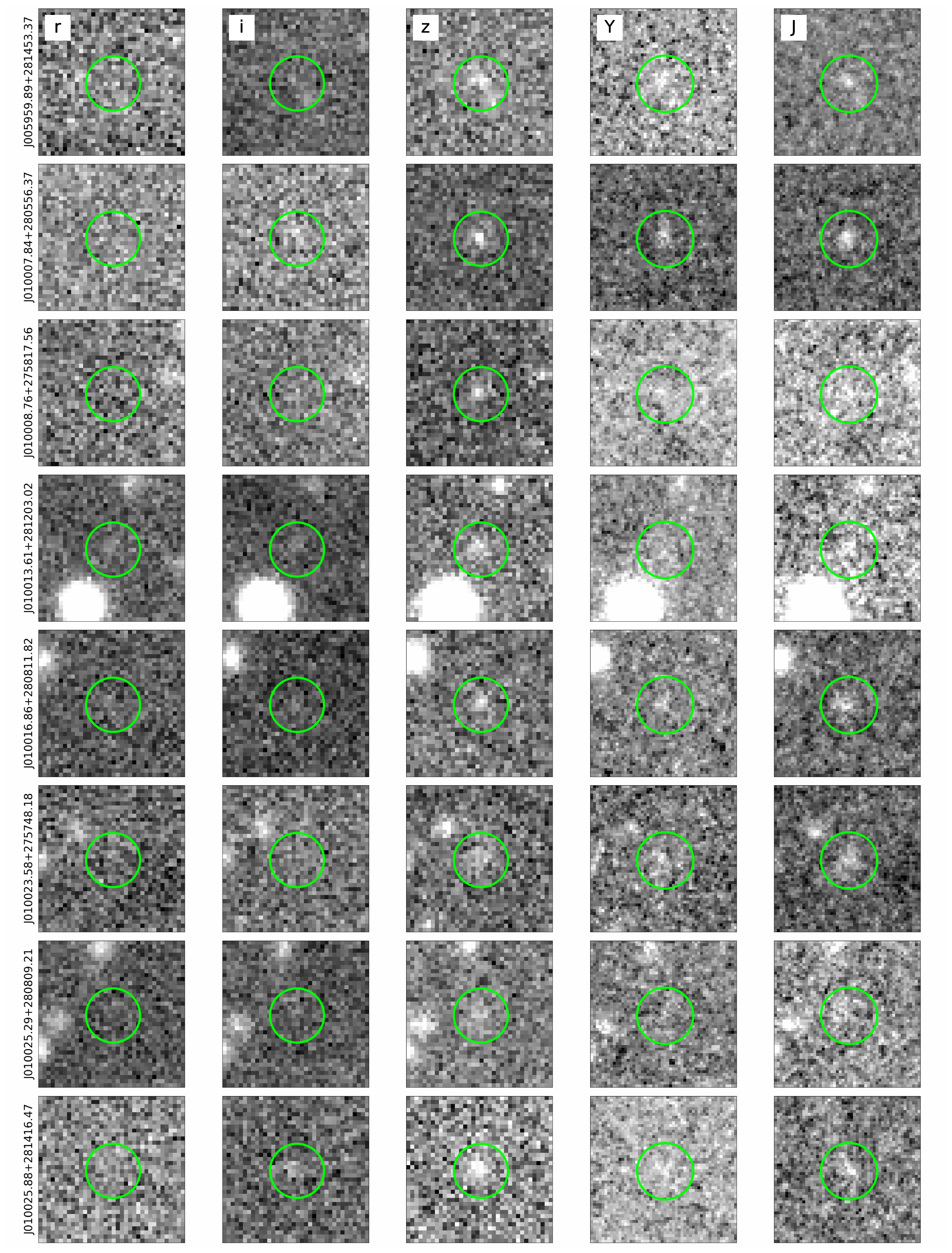}
    \caption{A continuation of Fig.~\ref{fig:cuts}.}
\end{figure}

\begin{figure}[ht]
    \centering
    \includegraphics[width=0.85\linewidth]{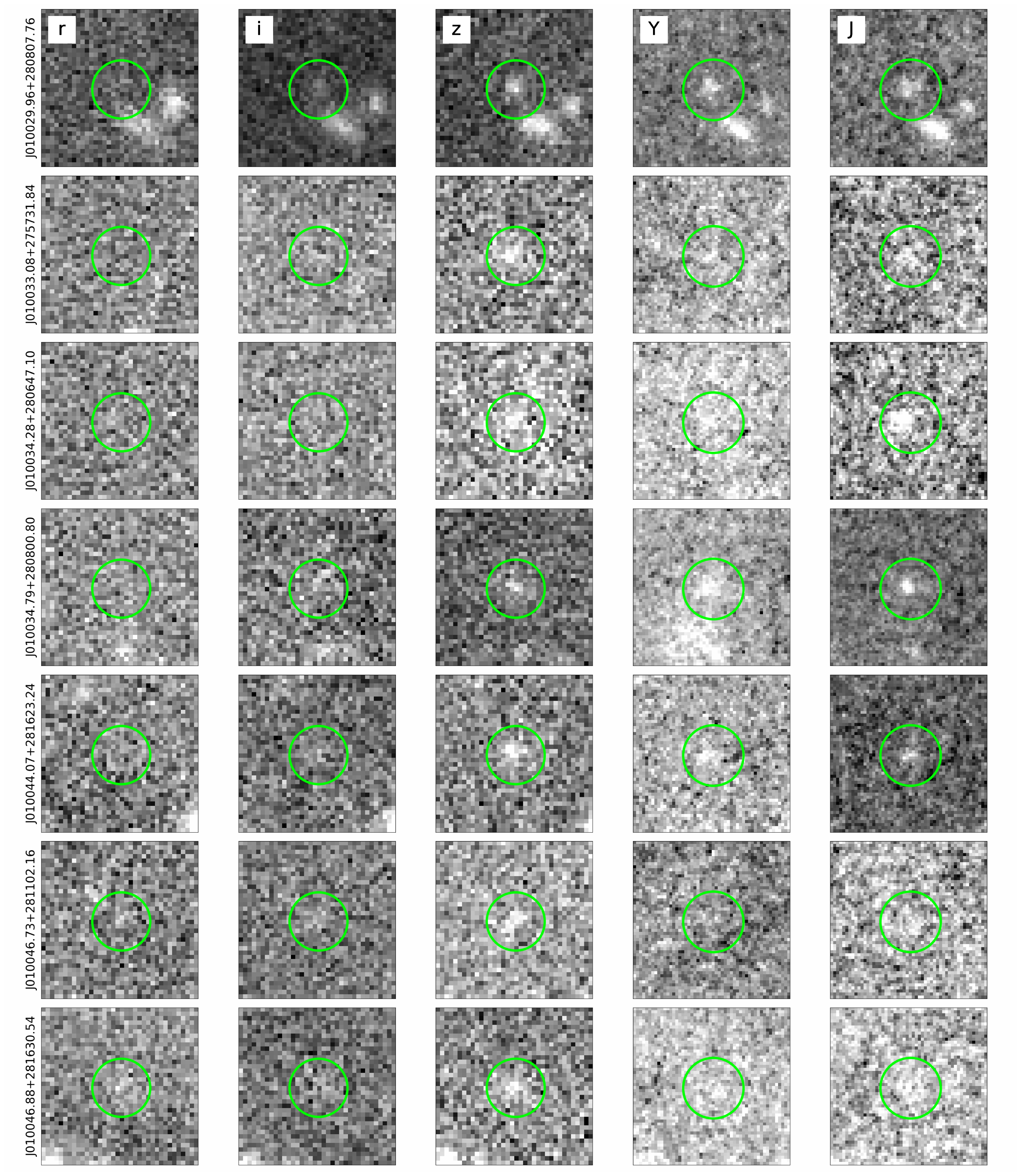}
    \caption{A continuation of Fig.~\ref{fig:cuts}.}
\end{figure}

\begin{figure}[ht]
    \centering
    \includegraphics[width=0.85\linewidth]{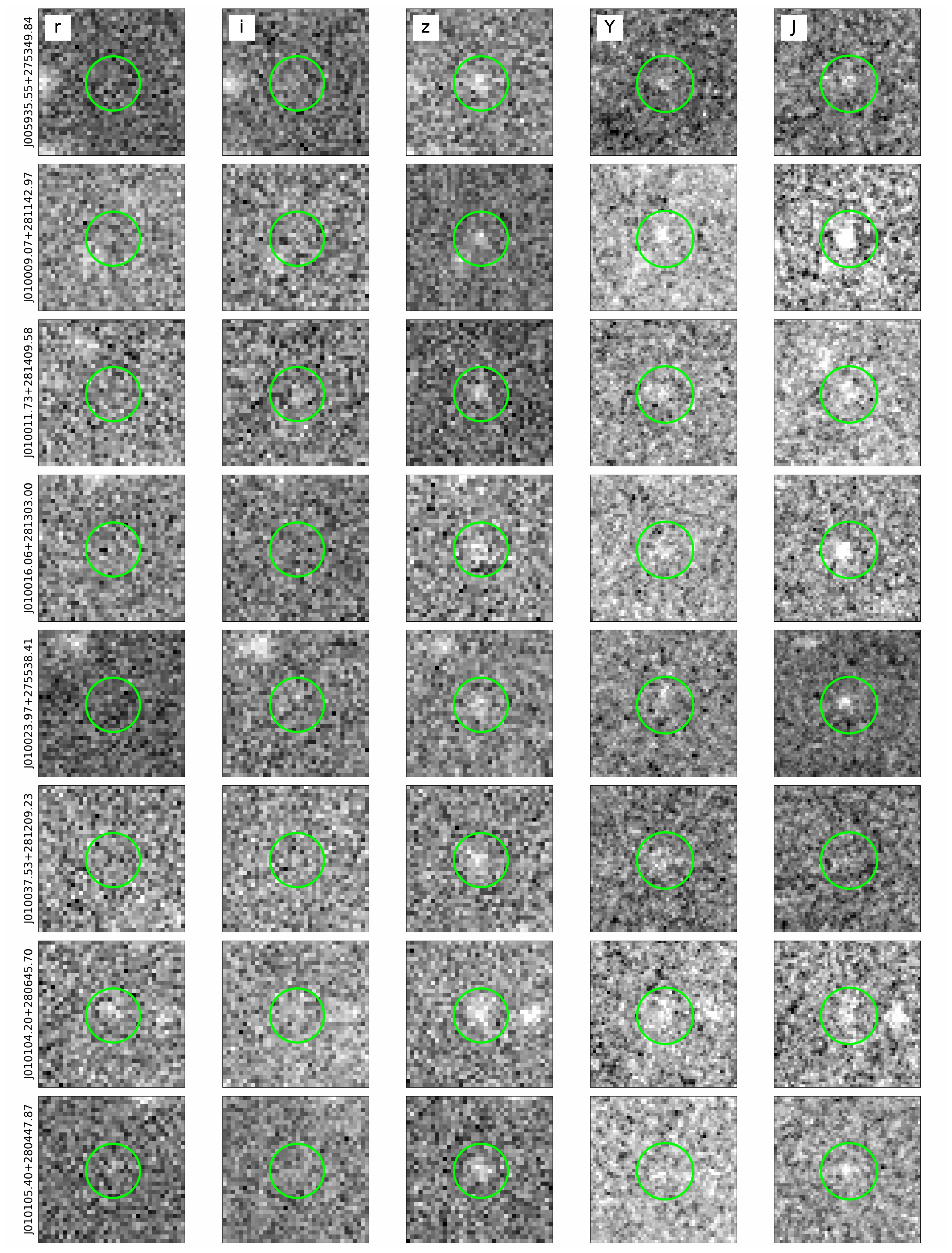}
    \caption{Continuation of Fig.~\ref{fig:cuts}, but showing the low-confidence targets listed in Table~\ref{props}.}
    \label{fig:enter-label}
\end{figure}
\clearpage
\bibliographystyle{aasjournal} 
\bibliography{refs}

\end{document}